# Predicting high dengue incidence in municipalities of Brazil using path signatures

**Daniel A.M. Villela**

Program of Scientific Computing, Oswaldo Cruz Foundation, Rio de Janeiro, Brazil

E-mail: daniel . villela@fiocruz.br

## Abstract

Predicting whether to expect a high incidence of infectious diseases is critical for health surveillance. In the epidemiology of dengue, environmental conditions can significantly impact the transmission of the virus. Utilizing epidemiological indicators alongside environmental variables can enhance predictions of dengue incidence risk. This study analyzed a dataset of weekly case numbers, temperature, and humidity across Brazilian municipalities to forecast the risk of high dengue incidence using data from 2014 to 2023. The framework involved constructing path signatures and applying lasso regression for binary outcomes. Sensitivity reached 75%, while specificity was extremely high, ranging from 75% to 100%. The best performance was observed with information gathered after 35 weeks of observations using data augmentation via embedding techniques. The use of path signatures effectively captures the stream of information given by epidemiological and climate variables that influence dengue transmission. This framework could be instrumental in optimizing resources to predict high dengue risk in municipalities in Brazil and other countries after learning these country patterns.

## 1 Introduction

Arbovirus infections, such as dengue, lead to millions of cases worldwide each year. In Brazil alone, approximately 1.7 million dengue cases were reported in 2023 [1]. According to the Ministry of Health in Brazil, dengue cases exceeded 1 million in 2015, 2016, 2019, 2022, and 2023 (source: tabnet.datasus.gov.br, data visualized in December 2024). In contrast, the incidence of dengue was significantly lower in other years, such as 2017 and 2018, even though cases were still reported in the hundreds. Accurate predictions of the risk of high incidence are vital for health preparedness.

Surveillance teams often seek answers regarding the expected number of cases for an upcoming season. Management in these teams can effectively allocate resources to control transmission. For instance, strategies such as conducting active surveys are known to enhance surveillance [2]. Outbreaks, especially for arbovirus infections, typically depend on epidemiological and environmental conditions, including climate factors [3]. Climate factors have been linked to expanding conditions conducive to dengue transmission in Brazil [4]. The dengue cycle in Brazil's urban centers is typically characterized by vectorial transmission with *Aedes aegypti* as the vector. Climate might impact vector abundance, survival, and incubation period of the virus, among other traits.

Path signatures are a robust framework for describing time series data sets [5]. This technique has been utilized for learning and distinguishing patterns [6]. Path signatures have successfully been applied in retrospective studies for diagnosing Alzheimer's disease [7]. While they have been used in other health studies [8,9], these applications typically focused on individual data. The primary task involved predicting individual outcomes based on a given set of features. However, path signatures have not yet been explored for predicting epidemiological outcomes using a set of features on population levels.

The concept involves using various epidemiological descriptors and other variables to help explain population trends. For instance, in the case of the dengue virus, data on several variables, such as the number of cases, temperature, and humidity, fluctuate over time. This information can be summarized into a sequence of numbers, referred to as a "signature," representing a particular disease spread trajectory. As more data is collected, a stream of information develops to describe this trajectory.

In this study, we develop a framework to analyze the streaming data of epidemiological and climate factors associated with dengue infections. We aim to evaluate the hypothesis that these path signatures can help predict whether a given season will have a high dengue incidence. We apply this framework to surveillance data of dengue cases reported in the capitals of Brazilian states covering 2014 to 2023.

## 2 Methodology

**Epidemiological and climate data**

The weekly number of dengue cases was obtained from the Brazilian National System of Notifications (SINAN) for all state capitals from 2014 to 2023. The time series data for each municipality was segmented into year-long periods, starting from epidemiological week 27 and extending to epidemiological week 26 of the following year. Each segment corresponds to a season, such as 2014/2015, 2015/2016, and so on, until 2023. In cities such as Rio de Janeiro, dengue transmission typically begins at the end of the year, with a decline during the first half of the following year. However, this cycle varies annually and might differ among municipalities. Brazil has 27 state capitals, resulting in a dataset of 270 year-long time series, each represented by a pair of municipalities and seasons. Each pair also includes a total cumulative number of cases, representing the seasonal count of dengue cases.

The notification data from SINAN was accessed through the Infodengue project (infodengue.mat.br) [10]. Additional variables, including weekly minimum temperature, average temperature, minimum humidity, and average humidity, were also obtained from the Infodengue project. The climate data in Infodengue is sourced from meteorological stations located within the respective municipalities [11].

The dataset was split into training and testing sets, with two-thirds of the time series used for training and one-third for testing. The division into training and testing was conducted through random sampling, adhering to these proportions.

**Outcomes of interest**

The mean and standard deviation were calculated for the logarithm of the cumulative number of cases. Each pair of municipality/year was classified based on whether its cumulative incidence fell within the highest number of seasonal cases, determined by a specified cutoff. This cutoff, expressed on a logarithmic scale, was derived from the mean value (logarithm) of cases plus the standard deviation, adjusted by a factor corresponding to a $k$ percentile. We evaluated different scenarios varying the value of $k$.

The highest incidences represent a portion of 1 - $k$ of the total set. For example, if $k$ is set at 80%, this would classify pairs so that approximately 80% fall below the cutoff, resulting in an outcome of 0. Conversely, the top 20% (1 - 0.80) incidences would have an outcome of 1, indicating that these are the 20% of units with the highest incidences.

**Construction of paths and path signatures**

For a given pair of municipality/season, a matrix of features $V_{n \times w}$, given by $n$ features and $w$ weeks describes a path. It is important to note that not necessarily all combinations of municipality and season have the same number of observations for these features. The features are on a weekly basis: the number of dengue cases, the cumulative number of dengue cases, average temperature, minimum temperature, average humidity, and minimum humidity. In a data augmentation step, we also enhance the feature set by incorporating the partial cumulative incidence. Many studies, such as those by Fermanian et al.[12], have highlighted the necessity of data embedding.

One evaluated option is to add the elapsed time in weeks to the dataset as an initial modification. Several studies have employed lead-lag transformations [5,12]. These techniques were applied to all the original variables in this study. Consequently, the embedding methods we utilized were categorized as "None," "Time," and "Time/Lead-Lag" for cumulative incidence, along with all three of these methods applied to the logarithmic value of cumulative incidence.

For each path, a signature is calculated from iterated integrals [13]:

$$S(X) = (1, \ \{S_{i_1}\}, \{S_{i_2}\}, \ldots, \{S_{i_n}\}, \{S_{i_1,i_1}\}, \{S_{i_1,i_2}\}, \ldots, \{S_{i_1,i_1,i_1}\}, \ldots),$$

where $S_{i_1,\ldots,i_n}(X) = \int \ldots \int_{1 \leq i_1 \leq \cdots \leq i_n} dX_{1_1} \ldots dX_{1_n}$.

Theoretically the signature captures the interactions between features at $l=1, 2, 3, \ldots \infty$ levels. For practical reasons, the signature is truncated at a maximum level $m$. Hence, terms on the fourth level such as $\{S_{i_1,i_1,i_1,i_1}\}$, $\{S_{i_1,i_1,i_1,i_2}\}$, etc. were included. The log-signature was used to reduce the number of terms and reduce redundancies [14] and the value used for truncation was $m=4$. For instance, with 7 features and $m=4$, this truncation leads to 728 terms in the log-signature.

**Statistical analysis**

The set of signatures $S_{c \times z}$, where $z$ is the length of a signature at the chosen truncation level and $c$ is the total number of combinations of municipality/season, is used as variables for regression analysis. Each pair given by municipality and season has an outcome indicating whether the season falls within a specific $k$ percentile of total incidence. For a vector $y$ representing outcomes for all municipality/season pairs, we analyze the data using logistic regression with lasso penalization [15], applying normalized terms, given means and standard deviations, from the signatures derived from the training data. This regression is combined with lasso regularization because it can efficiently reduce or eliminate the influence of irrelevant components due to a lack of association or statistical significance. This approach is critical given the length of the signatures and log-signatures. To predict outcomes, we use the best coefficients from the model, which minimizes the lasso hyperparameter $\lambda$, along with the normalized terms from the signature of the test data. The number of weeks of observation varied from 25 to 50 weeks for predicting with the testing set.

The total counts of positive and negative predictions, compared to the actual counts of positives and negatives from the testing data, form a contingency table. From these tables, we derive values for sensitivity and specificity. These values are calculated while varying the $k$ percentile, the number of weeks observed, and the chosen data augmentation method.

The implementation utilizes the Python package esig to obtain signatures and log-signatures, while the analysis tool was developed in R using the reticulate package. We used the glmNet and glmNetUtils packages for lasso regression.

## 3 Results

The analysis focused on 27 state capitals in Brazil over a series of 10-year periods. Out of these, 10 periods were excluded due to insufficient data from the original Infodengue dataset. The total study dataset was comprised of 260 pairs of municipalities and periods, which were used to examine the study variables.

Table 1 presents the distribution of variables related to dengue incidences and environmental factors. There was significant variability in incidences, as indicated by large standard deviations, despite a skewed distribution. Applying a logarithmic scale to the incidences helped to normalize the distribution. The coefficient of variation (ratio of standard deviation to average cumulative incidence) ranged from 0.16 in 2023 to 0.27 in 2017. The cumulative incidence was lowest in 2020 and highest in 2023, demonstrating substantial variability.

**Table 1.** Distribution of variables maximum weekly incidence, cumulative temperature (over the years), temperature (average and minimum), humidity (average and minimum). The year refers to the year of the start of the considered period.

| Variable (unit) | Year | Mean – absolute (St. Dev.) | Mean - $\log_{10}$ (St. dev.) | Max | Min |
|---|---|---|---|---|---|
| Max Incidence (weekly cases) | all years | 75.4 (142.6) | - | 1259.4 | 0.8 |
| Average temperature (Celsius) | all years | 24.2 (2.9) | - | 29.2 | 17.3 |
| Minimum temperature (Celsius) | all years | 21.6 (3.2) | - | 26.3 | 14.2 |
| Average Humidity (%) | all years | 78.4 (7.4) | - | 94.4 | 59.1 |
| Minimum humidity (%) | all years | 63.8 (10.7) | - | 83.5 | 33.7 |
| Cumulative incidence | 2014 | 855.0 (1087.5) | 2.7 (0.5) | 5261.5 | 42.0 |
| Cumulative incidence | 2015 | 1350.0 (1682.5) | 2.9 (0.5) | 7394.9 | 101.1 |
| Cumulative incidence | 2016 | 472.9 (597.6) | 2.4 (0.5) | 2400.6 | 48.4 |
| Cumulative incidence | 2017 | 319.0 (477.9) | 2.2 (0.5) | 2071.2 | 16.1 |
| Cumulative incidence | 2018 | 1084.9 (1069.2) | 2.6 (0.6) | 5846.5 | 44.5 |
| Cumulative incidence | 2019 | 668.4 (836.8) | 2.5 (0.6) | 3701.2 | 24.3 |
| Cumulative incidence | 2020 | 375.3 (566.5) | 2.2 (0.6) | 2803.0 | 5.9 |
| Cumulative incidence | 2021 | 1059.3 (1655.1) | 2.6 (0.6) | 7814.0 | 42.2 |
| Cumulative incidence | 2022 | 1373.0 (1954.2) | 2.8 (0.6) | 8698.9 | 35.3 |
| Cumulative incidence | 2023 | 2928.8 (3735.1) | 3.2 (0.5) | 14187.4 | 322.2 |

The timing of peak incidence also showed significant variation. The mean peak incidence occurred around week 34.6, with a median of 38 weeks. The observation range for peak incidence spanned from week 1 to week 53. Given that the reference starting week for each period is week 27, weeks 35 and 38 correspond to late February and mid-March, respectively.

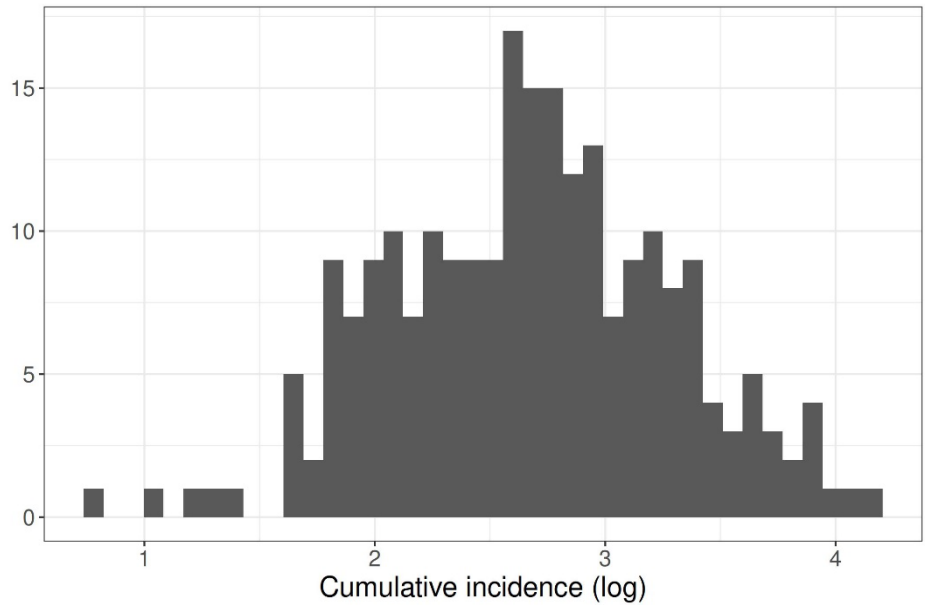

**Figure 1.** Distribution of cumulative incidence (log scale) over all pairs of period/municipalities considered.

Figure 1 shows that the cumulative incidence on the log scale is sufficiently close to a normal distribution (p=0.79, Shapiro-Wilk test). The mean value is 2.67, and the standard deviation is 0.61. These values were used to separate the observations with higher incidences according to different percentiles for training and test datasets.

The number of predictions with a positive rate exceeding 70% was higher when considering 46 to 50 weeks of observation (Figure 2). However, a significant number of predictions also surpassed the 70% threshold with 36 to 40 weeks of observation. Additionally, the performance of the embedding that included both time and lead-lag transformations was expected to be superior to other methods. The time-only embedding performed comparably well, showing only a slight decline in effectiveness. As expected, the sensitivity of the observations increases with more weeks of data collected.

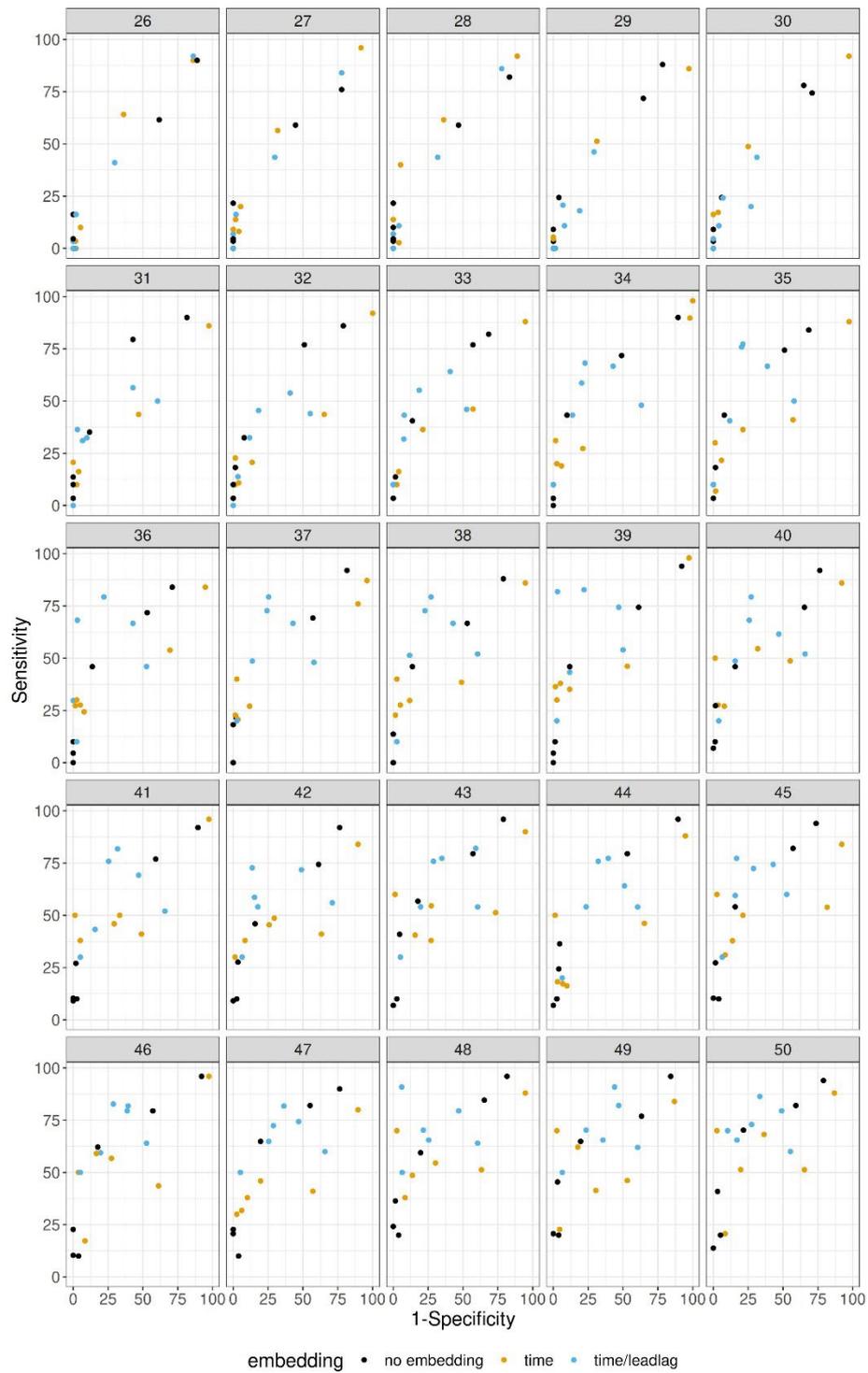

**Figure 2**. True Positive Rate against False Positive Rate for *k* from *k*=0.4 to 0.9 and time in testing dataset varying from week 25 to 50. The three types of embedding (colors) represent the evaluations when week number and a leadlag procedure were applied, only week number and no additional feature (only study variables).

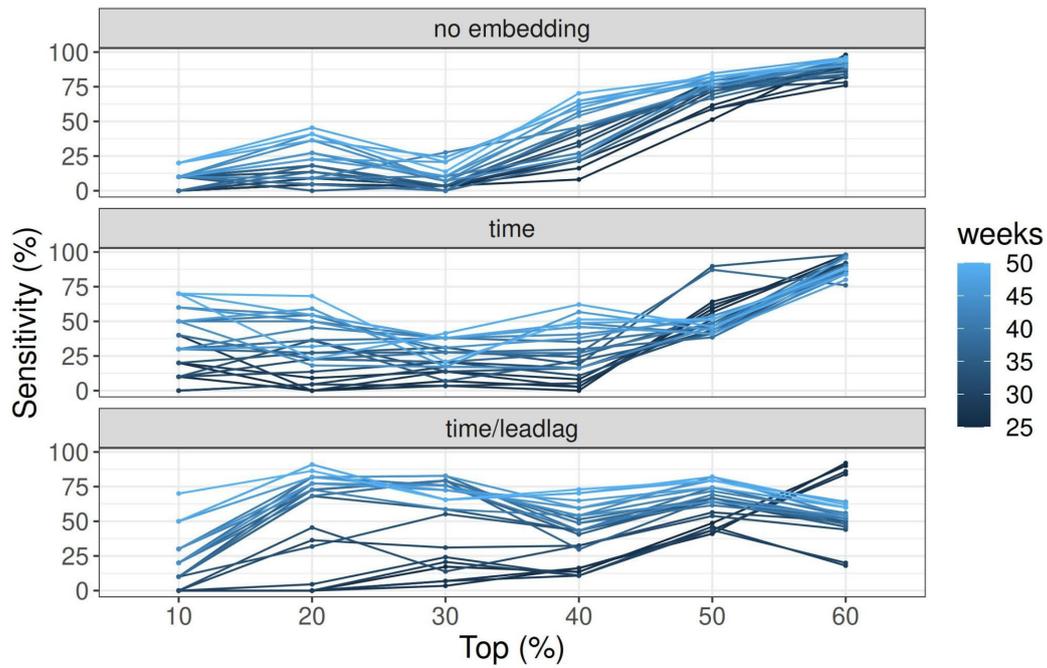

**Figure 3**. The sensitivity over the evaluated percentiles and presented by Top %. The evaluations covered percentiles 40% to 90% by increment of 10% (dots). Hence, the Top 10% to Top 60% highest incidences. The lines connect the evaluations using the same number of weeks used for prediction (color).

A true positive rate (sensitivity) of 75% was achievable, particularly when the number of weeks of observation was high and for the Top 20% and Top 30%. For most observations, sensitivity remained above 50% and showed relative stability across the tested percentile values. However, sensitivity was lower for the evaluations with Top 10%. The embedding that included time and lead-lag transformations demonstrated improved sensitivity for all percentiles, except when the observation period was close to 25 weeks. As the number of weeks of observation increased, sensitivity consistently improved, ultimately exceeding 75%.

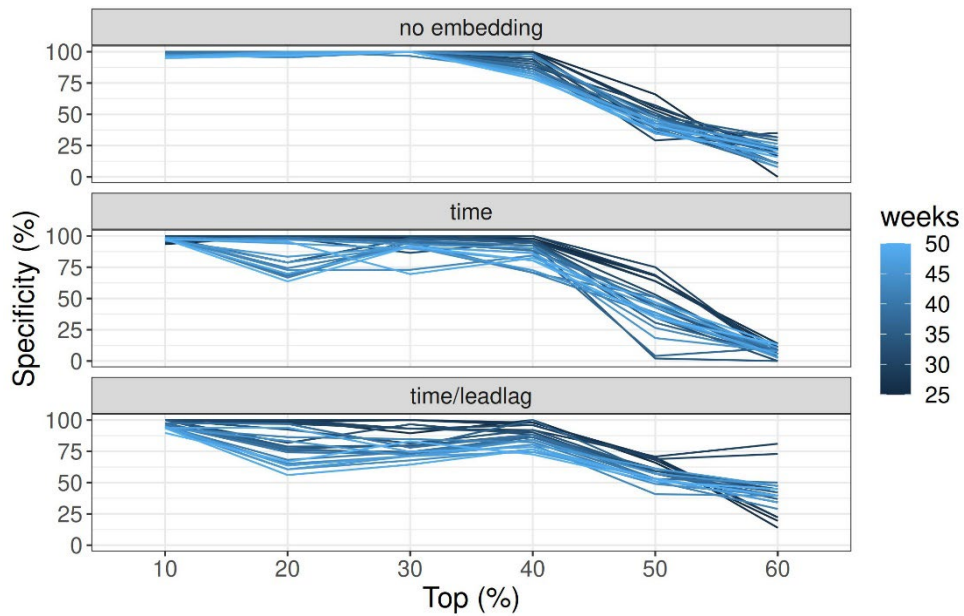

**Figure 4**. The specificity over the evaluated percentiles. The evaluations covered percentiles 40% to 90% by increment of 10% (dots) – Top 10% to Top 60%. The lines connect the evaluations using the same number of weeks used for prediction (color).

Specificity remained consistently above 75% for most tested scenarios, particularly within the top 10% to top 40% when using embedding methods (Figure 4). For the evaluations with Top 50% and Top 60%, the embedding that incorporated time and lead-lag transformations demonstrated better specificity. Additionally, the effect of the number of observation weeks starts close to 100% and gradually decreases to approximately 75% as the number of observation weeks increases.

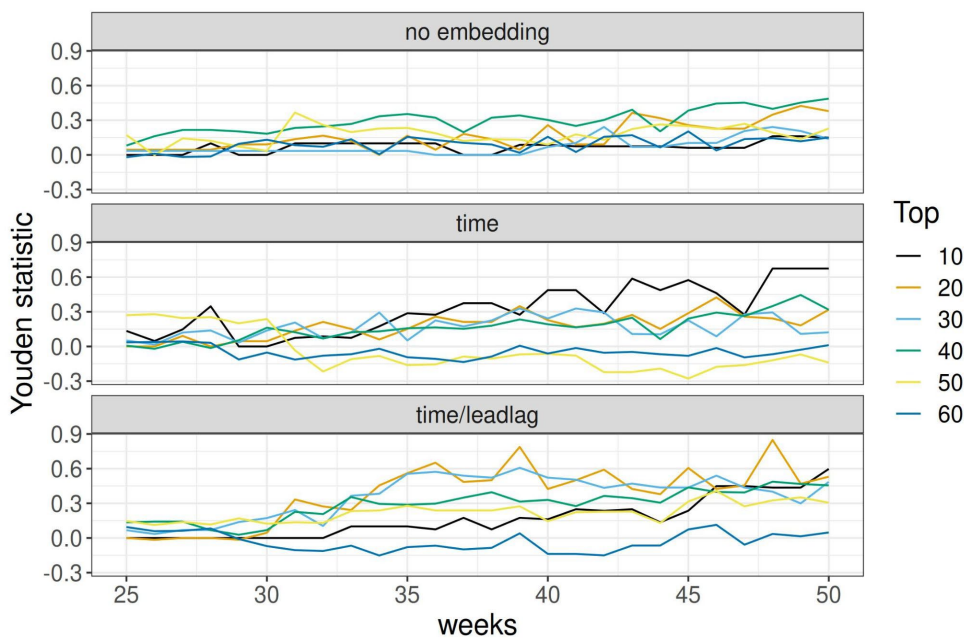

**Figure 5**. Youden statistic as the number of weeks used for prediction increased. Values can vary from -1 to 1, however, the scale was adapted for better visualization.

The Youden statistic demonstrates values exceeding 0.6 when applying the embedding with time and lead-lag transformation for the evaluation with the Top 20%, using 35 weeks of observations (Figure 5). In cases without embedding, the Youden statistic was positive in nearly all tested configurations. However, the Youden statistic did not improve as significantly with increased observations. Embedding only time, the performance improved with more observations, particularly for the evaluation with the Top 10% incidences.

## 4 Discussion

Preparedness for disease outbreaks allows for the effective allocation of efforts and resources to the areas that need them most, helping to prevent cases, particularly severe ones. During a potential dengue transmission season, as initial cases emerge in the first few weeks and climate-related indicators are collected, it becomes important to determine whether the current season will likely witness one of the highest incidences based on historical data. This study demonstrates good sensitivity and high specificity when predictions are made using path signatures. Various alternatives were evaluated within this framework, and sensitivity exceeded 75% even when considering outcomes in the Top 20% or Top 30% of highest incidences. While path signatures have been used in health studies with individual outcomes [7, 8, 9], this study is the first to apply the path signature approach for predicting epidemiological outcomes at the population level. The analysis focused on dengue transmission in Brazilian municipalities. The framework can be used for other arbovirus infections, other relevant infectious diseases, and municipalities of other countries once the respective databases are curated and the learning process is applied.

Even with more uneventful outcomes, such as the Top 10% incidences, when sensitivity is expected to decrease, performance remains reasonable when using embeddings that include both time and lead-lag transformations. The performance of time-only embeddings was slightly worse compared to those that included the lead-lag transformation. This was unexpected, as lead-lag transformations have demonstrated good performance in other settings [12] and can be computationally intensive.

Sensitivity was higher than 50% when at least 35 weeks of observations were taken into account in the testing dataset, arguably better than a random strategy, notably when municipalities reached their peak incidence on average. Sensitivity values can reach approximately 75%, while specificity rates often range from 75% to 100%. Such levels are advantageous for making predictions. The drawback is that high sensitivity values are typically associated with a larger number of weeks of observation.

The task of predicting outcomes in epidemiological settings remains challenging. The best performance, as indicated by the Youden statistic for the 80th percentile with embedding time and lead-lag analysis, shows that at least 35 weeks of observations are required to achieve this level of accuracy. Given that the mean peak of cases occurs around 36 weeks (with a median of 38 weeks), a surveillance team would need to wait several weeks before making a reliable decision. With only 25 weeks of observations, the true positive rate typically equals the false positive rate, making it essentially a random chance scenario. Additionally, the peak values of weekly incidence varied significantly across the study data and the timing of these peaks.

The concept of learning for epidemiological outcomes, particularly regarding dengue, has been studied using various techniques. Other works [20, 21] have applied machine learning methods to this issue. For instance, Hii et al. [17] utilized climate data and historical dengue case information to predict future cases using Poisson multivariate regression. Koplewits et al. [16] employed a combination of epidemiological variables and internet search data to generate nowcasts of dengue cases in Brazil, using lasso regression as part of their methodology. Similarly, Shi et al. [18] applied lasso regression with multiple indicators to forecast dengue cases in Singapore. Yamana et al. [19] used ensemble models to characterize indicators of dengue incidence, such as peak timing, based on data from Puerto Rico. While all these studies advance the field of dengue epidemiology prediction, they do not utilize summarizing techniques and often focus on slightly different outcomes, although related, such as forecasting or nowcasting cases.

The dataset size is important for evaluation, and while it was limited to a set of municipalities and features, the whole data collection was robust and sufficient for the study. The research utilized five original features over a span of 10 years across 27 municipalities. This provided a substantial collection of data, and for the statistical analysis, there were 260 pairs of municipality/year combinations. Additional climate factors could also be introduced in future studies. Still, the current analysis included four variables related to temperature and humidity, which are known to be associated with elements of arbovirus transmission [3]. The study's scope was constrained by the number of features available in the Infodengue project, which acts as an aggregator. Nonetheless, the variables we used were adequate for evaluation. Unfortunately, detailed climate variables at a micro-resolution are not accessible. However, one advantage of using state capitals regarding climate variables is that they often have nearby airports with regular climate data. State capitals typically have airports or are in close proximity to one another, which ensures better climate data availability given the presence of meteorological stations at airports. Also, the set of municipalities provides a diverse dataset, as some municipalities have populations in the hundreds of thousands, while others, like São Paulo, boast populations exceeding 10 million residents. Therefore, data availability was good and should accurately reflect the climate conditions in the municipalities. Another possible limiting factor is that the practical aspects of path signature analysis required truncation, and the number of interactions used for truncation generated large log-signatures, signaling a robust analysis.

The task of making predictions in epidemiological contexts is essential for effective preparedness. There is an increasing availability of data at epidemiological, socio-demographic, and environmental levels, including climate factors. This study demonstrated that utilizing path signatures of variables efficiently describes the data related to dengue incidence and climate variables. The application of lasso regression allows us to identify whether a municipality is likely to experience one of the highest incidence rates. Advancing this approach and these techniques will improve prediction performance in similar contexts, making efforts and resource use more efficient and ultimately helping to prevent cases in long-term scenarios.

## Acknowledgments

DAMV is grateful for the National Council for Scientific and Technological Development (CNPq/Brazil, Ref. 312282/2022-2), Fundação Carlos Chagas Filho de Amparo à Pesquisa do Estado do Rio de Janeiro (Ref. E-26/204.108/2024), and CAPES (Service Code 001). DAMV is grateful to the Center for Health and Wellbeing (CHW) at Princeton University, as part of this work was done during his time as Visiting Research Scholar at CHW.